\documentstyle[preprint,aps]{revtex}
\begin{document}
\def \beq{\begin{equation}}
\def \eeq{\end{equation}}
\def \beqarr{\begin{eqnarray}}
\def \eeqarr{\end{eqnarray}}

\draft

\title{Fractional Charge }
\author{R. Rajaraman}

\address{School of Physical Sciences, Jawaharlal Nehru University,
New Delhi 110067, India}

%\date{\today}

\maketitle

\begin{abstract}
The origin and quantum status of Fractional Charge in
polyacetylne and field theory are reviewed, along with
reminiscences of collaboration with John Bell on the subject.

\end{abstract}
\vskip 3.0 in

 [  Invited Lecture at the International Conference "Quantum (Un)speakables"
  in honour of John Bell, held at Vienna,  November  10-14 , 2000. ]  

\vfill
\eject

\section{Introduction}

I cannot claim the privilege of having known Professor John Bell
as intimately or for as long as some of the other participants in
this conference to honour his memory. But there was a period of
about 3 years from 1982 to 1985 when we interacted a fair amount
and conducted (mostly at long distance) a collaboration trying to
understand the then recently discovered phenomenon of fractional
charge. During this period we had the opportunity of hosting him
at my Institute in Bangalore. He came home for dinner and met my
family. I in turn went to CERN , first to write up our work and
later for a long visit, when I had the pleasure of meeting his
wife Dr. Mary Bell and was the beneficiary of their kindness in
many ways.

During this interaction I developed , like so many people before
me, great respect for Professor Bell, not just for his great
originality and prowess as a theoretical physicist, but equally
for the precision of thought and language that he demanded of
himself and of those who were fortunate enough to work with him.
John Bell was more than an intellectual force in physics--- his
was a moral presence. One can ill-afford the loss of such a moral
presence in today's world, and it is a great tragedy that he was
taken away from us when we continue to need him so much.

Long before I met John, I had of course heard of him and his many
important contributions. In fact the very first time I had to
study his work was in the early 'sixties, when I was a
foot-soldier in Professor Hans Bethe's army attacking the nuclear
matter problem. All of you know that John had worked on a wide
range of subjects, but not all might know that one of his early
interests was nuclear matter, on which he had co-authored a long
review article with E.J.Squires way back in 1961. A few years
after that as I gravitated towards particle physics and quantum
field theory I had to study in great detail  his landmark
discovery with Roman Jackiw of the axial current anomaly.

Despite such overlap of interests, I met him only  much later in
life when he spent a couple of weeks at  the Indian Institute of Science in
Bangalore where I was then working. A colleague introduced us
and he made the usual polite enquiry -- "What are you working on
these days ?" I thought that rather than inflict on him the
different problems I was working on at that time, I would
instead get his "take" on the  very interesting but puzzling
claims of fractionally charged states that had recently emerged.

As it turned out, he had not heard about these developments. "You
mean fractionally charged like quarks ? Isn't that by now old and
well established ? ", says John , a slight note of wariness
creeping into his voice. " No, no ",  I hasten to add, "I don't
mean quarks .  By fractionally charged I refer to states with
fractional eigenvalues of the Number Operator.  These have been
discovered not just theoretically in soliton states of field
theory models, but are claimed to be present in down to earth
systems like polymer chains . What I don't understand is how  a
real polymer made of some finite number of electrons could, no
matter how it  twists itself into a soliton configuration,  
carry a fractional number of electrons." This produced a flicker
of a smile from him and I knew I had caught his interest. "Would
you like to tell me about it ?", he asked. So I ushered him into
my office and  closed the door . This was a lovely opportunity ,
having trapped the distinguished John Bell in one's office with
all avenues of escape sealed off, to summarize for him $\it{ab \
initio}$ the curious phenomenon of fractional charge and get his
opinion.

Let me begin  the next section with a pedagogical expansion of
what I described to John in the fading light of that winter
afternoon in Bangalore.

\section{Fractional Charge in Field Theory}

The phenomenon of fractional charge was first discovered by Jackiw
and Rebbi in their pioneering work in 1976 in the context of
soliton states in quantum field theory \cite{JR}. Three years
after the Jackiw-Rebbi work, but quite independently and in an
entirely different context,  Su, Schrieffer and Heeger \cite{SSH},
argued that a similar phenomenon can occur for soliton states in
the long chain molecule ${\it trans}$-Polyacetylene. See also the
work by Rice \cite{Rice}. Subsequently Jackiw and Schrieffer
\cite{JS} wrote a paper drawing attention to the similarity
between these results discovered in entirely different areas of
physics.

Let us begin with the Jackiw-Rebbi work . They discussed the
phenomenon in both one- dimensional and
 three-dimensional models. Let me for simplicity discuss just the
 one-dimensional example, which contains all the essential ideas.
Consider in (1+1) dimensions a Fermi field $\Psi (x,t)$ coupled to
a scalar field $\Phi(x,t)$ through a Lagrangian density

\beq  {\cal L} \ \ = {\cal L}_B \ \ + \ \ {\cal L}_F  \label{L} \eeq
where,

\beq {\cal L}_B \ = \ \ \ {1 \over 2g^2} \bigg[ ( {
\partial\Phi \over \partial t} )^2 \ \  -  \ ( {
\partial\Phi \over \partial x} )^2 \  \ - (1/2)( \Phi^2 - 1)^2\bigg] \label {LB}
\eeq
 and
\beq {\cal L}_F \  = \  \ {\bar \Psi} \bigg(i \partial_{ \mu} \gamma^{ \mu}
  \ -  \   \ m \Phi(x,t) \bigg) \Psi \label{LF}  \eeq
This example corresponds to a quartic double well potential in
$\Phi$. It is straighforward to generalise the results to other
potentials $U(\Phi)$ with symmetric double wells. The phenomenon
of fractional charge occurs in the soliton sector of this system.
To a wider readership it may be helpful to recall what is meant by
the soliton sector and how fermi fields are treated in such
sectors (\cite{Raj}).

 The bosonic sub-system ${\cal L}_B$, in the absence of
 the  Fermi field, has as its field equation
 \beq \bigg({\partial^2  \over \partial t^2} \ - \ \nabla^2 \bigg)
 \phi \ - \ \phi \ + \ \phi^3 \ = \ 0 \eeq
 This has the following two lowest energy classical solutions :
 \beq \phi (x,t) = \pm 1  \label{vac} \eeq
 As is well known the presence of these two degenerate classical solutions
 indicates spontaneous breaking of the $\Phi \leftrightarrow -\Phi$ symmetry
 of the Lagrangian. Around each of these classical solutions a separate vacuum
  state and a whole tower of Fock
 states can be built corresponding to the two phases this system
 permits. We will call these the vacuum sectors.

 In addition this system also has two other static (time independent)
 solutions. Those are the topological soliton solutions  -- the so called kink
  solution and its reflected partner, the anti-kink, given respectively by
  \beq \phi_S (x)  \ = \  \pm  \ tanh ({x / \sqrt2}) \label{kink} \eeq
 As per the general theory of semi classical quantization
of quantum fields, one can build two other separate towers of
states, one around each of these soliton solutions. The
topological index  $ n \ = \ \phi (\infty) \ - \ \phi(-\infty)$
becomes a superselection quantum number upon quantization and
prevents any states from these soliton sectors  from decaying into
the vacuum sector.
 In short, we have four sectors of states for
this bosonic system : two are the vacuum sectors built around
$\phi = \pm 1$ and the other two are the soliton and antisoliton
sectors built around the kink and antikink solutions in eq (\ref{kink}). For
more details on soliton sectors of states and their properties see
(\cite {Raj}).

Now consider the full system in (\ref{L}) including the Fermi
field.
\subsection{Vacuum sector}
To leading order in $\hbar$  one can replace in each sector the
Bose field operator $\Phi$  occurring in the fermi field
lagrangian (\ref{LF})  by the corresponding classical solution .
Thus in the vacuum sector built around $\phi = 1 $, the fermi
lagrangian (\ref{LF}) reduces in leading order to

\beq {\cal L}_F \  = \  \ {\bar \Psi} \bigg(i \partial_{ \mu} \gamma^{ \mu}
  \ -  \   \ m  \bigg) \Psi   \eeq
 This is just the free Dirac system with mass $m$ discussed in textbooks.
 In order to contrast with what  later  happens
  in the soliton sector , let us  recall why
 the Number (Total charge)  operator has only integral eigenvalues for
 the free Dirac system.
  Let us denote by  $u_k (x)$ and $ {\tilde u}_k (x)$
  the  positive and negative energy spinorial  solutions of the Dirac equation
\beqarr ( -i \alpha \partial_x  \ + \ \beta m ) u_k (x) \ &=& \
E_k \ u_k (x)  \nonumber \\
 \bigg( -i \alpha \partial_x  \ + \ \beta m) {\tilde u}_k (x) \ &=&
 \ - \ E_k \  {\tilde u}_k (x)  \label{Deqn} \eeqarr
  where $E_k \  = + \sqrt{k^2 + m^2}$ and spinor indices have been
  suppressed.
   The Dirac matrices for this 1+1 dimensional system can be taken to be
  $\alpha = \sigma_2$ and $\beta = \sigma_1$.
The Dirac field is expanded  in terms of these solutions and the
 destruction operators $b_k$ and $d_k$ obeying
the usual anticommutation rules  . \beq \Psi (x,t) \ = \ \sum_k \
[ b_k u_k e^{-iE_{k}t} \ + \ d_k ^{\dagger} {\tilde u}_k
e^{iE_{k}t} ] \label{expan} \eeq
 The vacuum state in the $\phi = 1 $ sector is given by the
 conditions
\beq b_k |vac \rangle \ = \ d_k |vac \rangle \ \ = \ 0 \eeq with
all the bosonic oscillators being in the ground state.

 Note that the third Pauli matrix $\sigma_3$ acts as the charge conjugation
matrix. It anti commutes with the Dirac hamiltonian in
(\ref{Deqn}) and generates for every positive energy solution $u_k
(x)$ of energy $E_k$ the corresponding negative energy solution of
energy $-E_k$ \  :
 \beq \sigma_3 u_k (x) \ = \ {\tilde u}_k (x)  \eeq
Hence all modes of the expansion  ( \ref{expan}) come in pairs with
positive and negative energy. There are no zero energy solutions
in the free massive Dirac equation.

Finally consider the charge density operator (which is really
 the number density operator)
 \beq \rho(x,t) \ = \ {1 \over 2} \bigg[ \Psi^{\dagger} (x,t)  \ ,
 \Psi (x,t)  \bigg] \label{rho}\eeq
 This commutator form of $\rho$ is standard; it is designed to be regularised
and odd under charge conjugation . Inserting the mode expansion
(\ref{expan}) and using the orthonormality of the Dirac solutions
the total charge becomes

\beqarr  Q \ &\equiv& \ \int dx  \ \rho(x,t) \nonumber \\
              &=& {1 \over 2} \ \sum_k \ \bigg( \ [ b^{\dagger}_k  \ ,b_k \ ]
           \ + \ [ d_k , d^{\dagger}_k ] \bigg) \nonumber \\
          &=& \sum_k \ \bigg( \ ( b^{\dagger}_k  \ b_k \ - \ 1/2 )
           \ - \ ( d^{\dagger}_k  d_k  - \ 1/2 )  \bigg) \nonumber \\
           &=&  \sum_k \bigg( b^{\dagger}_k  \ b_k \
           \ - \  d^{\dagger}_k  d_k \bigg) \label{charge} \eeqarr
Notice that the half-integers cancel term by term because of the
existence of paired positive and energy modes. Since \beq
(b^{\dagger}_k  \ b_k)^2        \ = \  b^{\dagger}_k  \ b_k
 \ \ , \ \ \ \ \ \  (d^{\dagger}_k  \ d_k)^2 \ = \  d^{\dagger}_k  \ d_k  \eeq
these have eigenvalues of only 0 or 1. Hence the familiar result in the
vacuum sector that the charge operator has only integer eigenvalues.

\subsection{soliton sector}
 Let us repeat exactly the same steps in the soliton sector.
Now we have to substitute for the Bose field operator in  the
Dirac Lagrangian the classical kink function in eq (\ref{kink}).
 The corresponding Dirac equation now becomes

\beq ( -i \alpha \partial_x  \ + \ \beta \ m \ tanh {x \over
\sqrt2} \ ) \psi (x) \ = \ E \psi (x) \eeq For the two components
$\psi_{1,2}$ of the Dirac spinor, this yields the coupled
equations \beqarr ( \ - \partial_x \ + \ m \ tanh {x \over \sqrt2}
\ ) \psi_2 \
&=& \ E \ \psi_1 \nonumber \\
( \  \partial_x \ + \ m \ tanh {x \over \sqrt2} \ ) \psi_1 \ &=& \
E \ \psi_2 \label{Deqn2}\eeqarr This equation will also have a set
of positive energy solutions $\eta_k (x) $ with some associated
energy $E_k$. We need not find these solutions explicitly. But
since the charge conjugation matrix $\sigma_3$ again anticommutes
with the Dirac hamiltonian, we know that for every positive energy
solution $\eta_k (x) $ there will exist a negative energy solution
${\tilde \eta}_k (x) $ with energy $- E_k$. But now there is also
an unpaired zero-energy solution \beq
 \eta_0 \ = \ \pmatrix{ A \ exp \bigg( -m \int^x dy  \ tanh \
 (y/\sqrt{2})\bigg) \cr
 0\cr}
\eeq Such a normalisable solution to (16) exists because the
soliton
 function $tanh {x \over \sqrt2}$ which forms the background potential
 for the Dirac spinor tends to opposing limits $\pm 1$ as $x \rightarrow
 \pm \infty$. In infinite spatial volume this solution has no partner.
 It is self charge conjugate : $\sigma_3 \eta_0 \ = \ \eta_0$.

The mode expansion of the Fermi field operator now becomes

\beq \Psi (x,t) \ = \ \sum_{k \neq 0} \ [ b_k \eta_k (x)
e^{-iE_{k}t} \ + \ d_k ^{\dagger} {\tilde \eta}_k (x) e^{iE_{k}t}
] \  + \ a \eta_0 (x) \label{expan2}  \eeq where a is the
destruction operator for the zero-mode.

Unlike the vacuum sector built around $\phi=1$ which had a unique
ground state ("the vacuum" in that sector), in the soliton sector
there are two degenerate ground states because of the existence of
the fermionic zero-mode. They are $|sol\rangle$ and
$|\hat{sol}\rangle$ obeying \beq a |sol\rangle = b_k |sol\rangle =
d_k |sol\rangle = \ 0 \ \eeq and \beq |\hat{sol}\rangle \equiv \
a^{\dag}|sol\rangle \ \ ; \ \ a \ |\hat{sol}\rangle \ =
|sol\rangle \eeq These are the two basic quantum soliton states of
this system. They are energetically degenerate, but are
distinguishable by their Charge.

\beqarr  Q \ &\equiv& \  {1 \over 2} \int dx \bigg[ \Psi^{\dagger}
(x,t) \ ,
 \Psi (x,t)  \bigg] \nonumber \\
              &=& {1 \over 2} \ \sum_k \ \bigg( \ [ b^{\dagger}_k  \ ,b_k \ ]
           \ + \ [ d_k , d^{\dagger}_k ] \bigg) \ +  \ 1/2 [ a^{\dag} \ , \ a \ ] \nonumber \\
          &=& \sum_k \ \bigg( \ ( b^{\dagger}_k  \ b_k \ - \ 1/2 )
           \ - \ ( d^{\dagger}_k  d_k  - \ 1/2 )  \bigg)  \ + \ (a^{\dag} a  \ - \ 1/2)
           \nonumber \\
           &=&  \sum_k \bigg( b^{\dagger}_k  \ b_k \
           \ - \  d^{\dagger}_k  d_k \bigg)  \ + \ a^{\dag} a  \ - \ 1/2 \label{charge2} \eeqarr

Notice that the piece (-1/2) coming from the zero- mode commutator
remains uncancelled because it does not have a charge conjugate
partner. Since the operators $b^{\dagger}_k \ b_k , \ \
d^{\dagger}_k \ d_k$ and $a^{\dag} a$ all have eigenvalues of 0 or
1, it then follows that the total Charge (Number) operator $Q$ has
half-integral eigenvalues . It should be emphasized that
eq(\ref{charge2}) is an operator equation for $Q$. The
half-integer appearing in it will be reflected in  its
\underline{eigenvalues} and not just its expectation values. In
particular, the two degenerate soliton states have  eigenvalues of
$\pm 1/2$ respectively for  the total number operator $Q$  :
 \beq Q \ |sol\rangle = -(1/2) |sol\rangle
\ \ \ \ \; \ \ \ \ Q \ |\hat{sol}\rangle =  (1/2) \
|\hat{sol}\rangle \eeq

This, in its barest form, is the prototype example of the
Jackiw-Rebbi discovery that in some field theories there can be
states  carrying fractional eigenvalues of the Number operator.
For  a fuller discussion including similar results in
three-dimensional models see their original paper \cite{JR}.

\section{Polyacetylene}

This discovery by Jackiw and Rebbi  was clearly  very remarkable.
But it is not  easy to "understand"  the result physically in
terms of our familiar intuition with quantum field theory. How can
the Number Operator which is widely used to count the number of
particles in QED and so many other theories, and which even in the
Higgs model used by Jackiw and Rebbi has only integer eigenvalues
in the vacuum sector, yield fractional values in another sector ?
Yet the proof given is so simple and transparent that, stare at it
as we may, we have no choice but to accept the result within the
parameters of its derivation. In my own attempts to make peace
 with the result (before embarking on the more detailed study with John),
I  loosely attributed  it to the vagaries of the infinite degrees
of freedom of continuum field theory.  [ We know from high school
that formally summing divergent series of integers can yield
fractions.] But clearly some further clarification of this
phenomenon was called for.

The need for clarification became more compelling when a similar
result was derived not in a model field theory, but in a
down-to-earth experimentally accessible polymer system by Su
Schrieffer and Heeger \cite{SSH}. Three years after the
Jackiw-Rebbi work, but quite independently they showed that the
same phenomenon of fermion number 1/2 can occur for soliton states
in the long chain molecule ${\it trans}$-Polyacetylene.

That the same phenomenon occurs in Polyacetylene (we will
henceforth drop the prefix "trans")  as in the Jackiw-Rebbi  field
theory is not accidental. As pointed out by Jackiw and Schrieffer
\cite{JS} , the former system has the same structure
 in the continuum limit as the field theory model. Polyacetylene
is the molecule $(CH)_n$ with large n, where the Carbon ions form
a long chain with the H atoms sticking out transverse to the
chain. If we consider the flow of electrons along the chain, the
system can be viewed as consisting of electrons and bosons (the
phonons of lattice vibration of the Carbon ions along the chain)
in one space dimension just as in the field theory model .

The Hamiltonian for 
the electron-lattice system can be taken,
 for   each  of   the two spin states of the electron and for
small ion displacements as (see \cite{JS}) :

\beq H \ \ = \ \sum_n \bigg[ \bigg( {p_n^{2} \over 2\mu} \ +
\ K/2 \ (u_{n+1}-u_n)^2 \bigg)  \\
 \ + \ \bigg( D^{\dag}_{n+1} D_n  \ + \ h.c. \bigg) \bigg(u_{n+1}-u_n \ - \
1/(2a) \ \bigg) \bigg] \label{polyHam} \eeq
 where $n$ labels lattice sites, $p_n$ and $u_n$ are ion momenta
and displacements respectively , $D_n$ is electron destruction
operator and $K$ and $a$  are constants. All other constants have
been absorbed into definitions for simplicity.

Our primary interest is in the fermions . As far as the lattice
vibrations go, let us just state the Su ${\it et \ al}$ result
\cite {SSH} that the lattice system undergoes dimerisation
doubling its spatial period (a Peierls transition) . There are two
degenerate ground states. In one ,  the mean value of the
displacement $u_n$ instead of being zero takes the staggered value
of \beq \phi_n \equiv \ 4(-1)^n u_n \ = \ 1 \eeq for all $n$. In
the other
  \beq \phi_n \ = \ -1 \eeq  The two degenerate
ground states  further provide the possibility of domain wall
(soliton) configurations connecting the two phases , i.e. $\phi_n
\rightarrow \pm 1   \ ( or \ \mp 1 ) $, as $n \rightarrow \pm
\infty $ respectively. Thus the boson subsystem has four sectors
of states -- two of them being soliton sectors -- just as in the
field theory discussed in the previous section.

Furthermore, the boson coordinates $u_n$ again act as the
background potential for the electrons . In fact the electronic
part of the hamiltonian (\ref{polyHam}) becomes in the continuum
approximation exactly the same as that of the Dirac field theory.
John and I , as part of our work on these systems \cite{BR},
offered a simple derivation showing that the polyacetylene
hamiltonian is just a realisation of the Kogut-Susskind lattice
regularisation of the Dirac system \cite{Kogut}. Let me sketch
that derivation. The electronic part of (\ref{polyHam}) is \beq
H_{elec} \ =  \ \ \sum_n \bigg[ D^{\dag}_{n+1} D_n  \ (u_{n+1}-u_n
\ - \ 1/(2a)  \ ) \bigg] \ + herm.conj \eeq Define staggered
variables \beq B_{2r-1} \equiv (-1)^r \ D_{2r-1} \ ; \ C_{2r}
\equiv (-1)^r \ D_{2r} \eeq with $\phi_n \equiv \ 4(-1)^n u_n $
for all $n$ as already defined above. Then, \beqarr H_{elec} \ &=&
\ \sum_r \bigg[\bigg( D^{\dag}_{2r+1} D_{2r}  \
  (u_{2r+1}-u_{2r} \ - \ 1/2a) \ \ + \ \  D^{\dag}_{2r} D_{2r-1}
    \  (u_{2r}-u_{2r-1} \ - \ 1/2a)\bigg)  \ + \  h.c. \bigg]  \nonumber \\
    &=& (1/4) \ \sum_r \  \bigg[  B^{\dag}_{2r+1} C_{2r}  \
  (\phi_{2r+1} + \phi_{2r} \ + \ 2/a) \ \ + \ \  C^{\dag}_{2r} B_{2r-1}
    \  (\phi_{2r}-\phi_{2r-1} \ - \ 2/a) \bigg] \ + \  h.c.  \nonumber \\
  &=& \sum_r   \bigg( {B^{\dag}_{2r+1} - B^{\dag}_{2r-1} \over 2a} \bigg)
  C_{2r} \ + \ (1/4)  B^{\dag}_{2r+1}   C_{2r} \ (\phi_{2r+1} + \phi_{2r})
\ \ +  \ (1/4)  B^{\dag}_{2r-1}   C_{2r} \ (\phi_{2r} + \phi_{2r-1}) 
\nonumber\\
 &+& \ \  h.c.  \label{Hpoly}\eeqarr

In the continuum limit, as $a \rightarrow 0$ and $ \sum_r
\rightarrow \int dx$,
 this reduces to
 \beqarr H_{elec} \ &=& \
 \int  \ dx \ \bigg( {\partial B^{\dag} \over \partial x} C \
 + \ C^{\dag} {\partial  \ B \over \partial x}   \bigg) \ + \ \phi (x) \
 (  B^{\dag} C + C^{\dag}B \ )  \nonumber \\
 &=& \int  \ dx \ \bigg( - \  B^{\dag}{\partial C  \over \partial x}  \
 + \ C^{\dag} {\partial  B \over \partial x}  \bigg) \ + \ \phi (x) \
 (  B^{\dag} C + C^{\dag}B \ ) \eeqarr
 This is just the Dirac Hamiltonian with Yukawa coupling :
 \beq H_F \ = \ \Psi^{\dag} \bigg( -i \sigma_2 {\partial \over dx} \ + \
 \sigma_1 \phi (x) \bigg) \Psi \eeq
 with the upper and lower components of the 2-spinor being identified
 respectively with the  odd-and even- site electron operators :
 \beq \Psi (x) \ \equiv \pmatrix{ B(x)\cr C(x) \cr } \eeq

 The charge operator operator at each site  also has the same
 commutator form  as in the continuum theory's charge density :
 \beq \rho_n \ = \  D^{\dag}_n  D_n - (1/2) \ = \
 (1/2) \bigg[ D^{\dag}_n , D_n \ \bigg] \eeq
 where the $1/2$ subtracted at each site can be attributed to the neutralizing
 background (ionic) charge $\it{per  \ spin  \ state}$.

 Given this mapping from the polyacetylene system to the field theory model
  used
 by Jackiw and Rebbi, it is not surprising that fractional charged states
 arise in the former system too.

 But in the context of a real polymer the result becomes still more mysterious.
 A molecule of Polyacetylene has after all some finite number
of electrons and it is hard to imagine how , even if it twists
itself into some topological soliton state, it could have a
fractional number of electrons. The need to understand this better
becameall the  more compelling because  Su, Schrieffer- and Heeger
referred to experimental signals supporting the result .
 See for instance ref (\cite{expt}). [ Because of the presence of the two
spin degrees of freedom in actual polyacetylene, as distinct from
a truly 1+1 dimensional  model, the fractional charge gets doubled
and becomes integral; but a signature of the effect can still be
observed in the form of an unfamiliar charge-spin combination of
excitations.  There will be neutral excitations with spin and
charged excitation which are spinless.]

It should be emphasised that an $\underline{expectation  \ value}$
 of one-half would cause no surprise . Obviously that can easily arise for an
 operator with integral eigenvalues. For instance if we take any particle
  in the ground state of a double-well potential and look for it in
 one of the wells, you will find it there half the times and not find it
 there the other half, giving an expectation value of 1/2.
 While in the Su et al paper, fractional values were not
 claimed as eigenvalues, in the prototype field theoretic case we have explicitly seen
 in sec.II  that the eigenvalues themselves are half-integers.
 This is what puzzled me and I was fortunate in having
 John Bell  join me in worrying about it further.

\section{Eigenvalue or Expectation value ?}

Now let me summarize what John  and I did in our attempt to
understand better what these half integral values of the number
operator mean \cite{RB}, \cite{BR} . [ We later learnt that
Kivelson and Schrieffer had also been independently investigating
the eigenvalue status of fractional charge around the same time
with essentially a similar resolution of the issues
 \cite{KS}. See also the work of Jackiw $\it{et al}$ which
provided further clarification \cite {J et al}. I will describe
here only the work done with John Bell.] Our strategy was to treat
the infrared and ultraviolet limits of the problem more carefully.
In one paper \cite{RB}, we looked at the continuum field theory
used by Jackiw and Rebbi, but in a finite volume, which could be
later taken to infinity. In a subsequent paper we considered what
would happen if we put an ultraviolet cut-off as well, by going to
 the finite polyacetylene chain\cite{BR}.

 Let me begin with the field theory example of
Jackiw and Rebbi which was discussed in section II. The result
derived there that the total charge operator has half-integral
eigenvalues cannot be disputed in the infinite spatial volume case
($L \ \rightarrow \infty$ in the one dimensional field theory
model considered there). However suppose we kept the spatial
volume 2L finite to start with, and later go the $L
 \rightarrow \infty$ limit \cite{RB}. In particular , in the soliton sector,
 we go back and solve the same Dirac equation
\beqarr ( \ - \partial_x \ + \ m \tanh {x \over \sqrt2} \ ) \psi_2
\
&=& \ E \ \psi_1 \nonumber \\
( \  \partial_x \ + \ m \ tanh {x \over \sqrt2} \ ) \psi_1 \ &=& \
E \ \psi_2 \label{Deqn2}\eeqarr but in a finite volume $ (-L \leq
x \leq L) $ with appropriate boundary conditions on the spinor
$\psi_\alpha $ at $x = \pm L$. The natural boundary conditions
which leave the Dirac hamiltonian hermitian are $\psi_{1,2} (-L) \
= \ \psi_{1,2}(L) $. Once again every positive  energy solution
$\eta_k$ will have a negative energy partner $\tilde {\eta}_k$. We
do  not need their details. But now there are $\underline{two}$
zero energy solutions :

\beq
 \eta_0 (x) \ = \ \pmatrix{ A \ exp \bigg( -m \int^{x}_0 dy  \ tanh \
 (y/\sqrt{2})\bigg) \cr  0\cr} \label{zero1} \eeq
and \beq
 \tilde{\eta}_0  (x) \ = \ \pmatrix{0\cr A \ exp \bigg( -mL \ +
 m \int^{x}_0  dy  \ tanh \
 (y/\sqrt{2})\bigg) \cr}   \label{zero2} \eeq
While the solution (\ref{zero1}) is the same one as before ,
localised around the origin, the second solution (\ref{zero2}) has
support near the edges $\pm L$. In section II where  we started
with an infinite volume problem, this second solution could not be
entertained. But for any finite L however large this second
solution, finite everywhere and normalisable,  is certainly
present. Each of these two zero-modes is charge self-conjugate and
both will appear in the expansion of the Dirac field operator,
along with the non-zero energy modes : \beq \Psi (x,t) \ = \
\sum_{k \neq 0} \ [ b_k \eta_k (x) e^{-iE_{k}t} \ + \ d_k
^{\dagger} {\tilde \eta}_k (x) e^{iE_{k}t} ] \  + \ a \eta_0 (x) \
+ \ c^{\dag} \ \tilde{\eta}_0 (x) \label{expan3}  \eeq
Correspondingly there will now be four degenerate ground states in
the soliton sector, namely, $|sol\rangle$  defined by \beq  b_k
|sol\rangle  \ = \ d_k |sol\rangle \ = \ a |sol\rangle\  \ = \ c
|sol\rangle \ = \ 0 \eeq and $ \widetilde{|sol\rangle} \ = \
a^{\dag} |sol\rangle \ , \  \overline{|sol\rangle} \ = \ c^{\dag}
|sol\rangle ,$ and $ \widetilde{|sol '\rangle} \ = \ a^{\dag}
c^{\dag}|sol\rangle $.

The expansion (\ref{expan3}) when inserted into the definition of
the charge operator \beq Q \ \equiv \ {1 \over 2} \int_{-L}^{L} dx
\bigg[ \Psi^{\dagger} (x,t) \ ,
 \Psi (x,t)  \bigg]\eeq will yield
 \beq Q \ = \ \sum_k \ \bigg( \ ( b^{\dagger}_k  \ b_k \ - \ 1/2 )
           \ - \ ( d^{\dagger}_k  d_k  - \ 1/2 )  \bigg)  \ + \ (a^{\dag} a  \ - \ 1/2)
 \ + \ (c^{\dag} c  \ - \ 1/2) \label{Qtot} \eeq
 We see that there are now no unpaired terms of 1/2 and the total charge
 operator has only integral eigenvalues.
One can now let the volume 2L become larger and larger and at each
stage both zero modes will be present and the total charge will
have only integral eigenvalues.

 That one can restore integral valued charges by this
  strategy of starting with a finite volume and then
letting $2L \rightarrow \infty $ can also be  understood in a
different way. Note that if we had closed our line $ (-L \leq x
\geq L) $ by identifying the points $x = \pm L$, that  is
equivalent to putting the system on a circle with a soliton at
 x=0 $ \underline{and} $ an antisoliton at x = L.
This  in turn amounts  to working in the $\underline{vacuum \
sector}$  , where we saw that the charge is integral.

While integral charge has been thus restored, if we were to stop
here we would have lost all the important physics unearthed by
Jackiw and Rebbi. Notice that while there are now two zero modes,
both localised with width of order $1/m$, one of them, ${\tilde
\eta}_0 (x)$ given in eq.(\ref{zero2}) is stuck at the edges of
the sample. But the other zero mode $\eta_0 (x)$ in (\ref{zero1}),
is in the middle and can move about as an excitation,  carrying
charge 1/2 in some sense that needs to be made more precise.

To pick just the charge carried by this middle zero-mode, we can
try and define a "partial charge operator" : \beq P_l \equiv \
\int_{-l}^{l}dx  \ \rho(x) \label{PCh}\eeq where $l, L \rightarrow
\infty$ in the order $L >>l >> 1/m$. In this limit the partial
charge $P_l$ clearly includes the central zero mode while
excluding the second zero mode at the edges $\pm L$. However, the
soliton ground state will not
 be an eigenstate of $ P_l$, but only of the total charge operator Q in
eq(\ref{Qtot}). The charge density operator $\rho(x)$ given in
eq(\ref{rho}) and  partial charge operators such as $ P_l$ will
excite particle-hole pairs when acting on the soliton state
$|sol\rangle.$  But consider expectation values in this soliton
state. \beqarr \langle \ sol| \ \rho(x)  \ |sol\rangle \ &=& \
(-1/2)\sum_{k \neq 0} \bigg[ \eta^{*}_k (x) \ \eta_k (x) \ - \
{\tilde \eta}_k
*(x){\tilde \eta}_k (x) \bigg] \nonumber \\
 &+& \ (1/2)\bigg({\tilde \eta}_0 (x){\tilde \eta}_0 (x) \ - \
 \eta_0
(x) \eta_0 (x) \bigg)
\nonumber \\
&=& (1/2)\bigg({\tilde \eta}_0 (x){\tilde \eta}_0 (x) \ - \ \eta_0
(x) \eta_0 (x) \bigg) \eeqarr where in the last line,
contributions from non-zero modes cancel for each k by charge
conjugation. Hence \beqarr \langle \ sol| \ P_l  \ |sol\rangle
 \ &=& \ \int^l_l \ dx (1/2)\bigg({\tilde \eta}^2_0 (x) \ - \
 \eta^2_0
(x)  \bigg) \ \nonumber \\
&=& -(1/2) \ \int^l_l \ dx  \eta^2_0(x) = \ -(1/2) \eeqarr
Because this is not an eigenstate of $P_l$ there will be
fluctuations. Since \beq ( P_l  \ - \ <P_l> ) |sol \rangle \ = \
\sum_{k,k'}  \ \int_l^l dx  \ ( \eta_k(x) \ \tilde \eta_{k'}(x)) \
| k, k' \rangle \eeq We have  \beq  \langle
sol | \  ( P_l  \ - \ <P_l> )^2  \ |sol \rangle \ = \  \sum_{k,
k'} | \ \int_l^l  dx \ \eta_k (x) \ \tilde \eta_{k'}(x) \ | \ ^2
\label{fluc}\eeq The integrand is positive and hence the
fluctuations are non-zero. In fact they diverge logarithmically.

Such large fluctuations in partial charges  are not special to
soliton states. They occur even in familiar vacuum states and are
because of the sharp boundary of the defining domain. One can kill
the fluctuations while still retaining a value of 1/2 by defining
a more sophisticated partial charge operator, one with fuzzy
edges. Let \beq \tilde{P}_{l,d} \ \equiv {1 \over d}
\int_{l}^{l+d}  \ dl' \ P_{l'} \label{fuzzy}\eeq When $L,l,d \
\rightarrow \infty$ with $L >> l, d$ then once again, \beq \langle
\ sol| \ \tilde{P}_{l,d} |sol\rangle \ = \ -(1/2) \eeq To obtain
the fluctuations of this operator note that $ [ \rho(x) \ - \
<\rho(x)> ]$ connects the state $|sol \rangle$ to the
particle-hole states $|k,k' \rangle$ . We therefore have

\beqarr \langle k,k'| P_l \ - \ <P_l>  |sol \rangle \ &=& \
\langle k,k'|
\int^l_{-l} dx  \ (\rho \ - \ <\rho> ) \   |sol \rangle \nonumber \\
&=& { 1 \over i(E_k \ + \ E_{k'})} \langle k,k'| \int^l_{-l} \ dx
{\partial \rho \over dt}   |sol \rangle \nonumber \\
&=& \ { 1 \over i(E_k \ + \ E_{k'})} \langle k,k'| \ ( j(l) \ - \
j(-l)) \  |sol \rangle \label{mael} \eeqarr by current
conservation. Note that the current $j$ has matrix elements  $
\langle k,k'| j(x) |sol \rangle \  \ = \ \eta^{*}_{k} (x)i \alpha
{\tilde \eta}_{k'}(x) $.  Hence, \beq \langle k,k'|
\tilde{P}_{l,d} \ - \ < \tilde{P}_{l,d}> |sol \rangle \ = \ {1
\over d} \int_{l}^{l+d} \ dl' \ { 1 \over i(E_k \ + \ E_{k'})}
\bigg[ \eta^{*}_{k} i \alpha {\tilde \eta _{k'}}\bigg]_{-l'}^{l'}
\eeq

 To obtain the exact result for this one must know
analytic expressions for all the non-zero energy modes $\eta_k
(x)$ and ${\tilde \eta_k} (x)$. We had in fact obtained these
solutions for the simplified case where the soliton function is
taken as a step function $\phi(x) \ = \ \Theta(x)$ instead of
$tanh \ x$. This simplification does not affect the issues of
interest to us. But even quite generally one can see that for
large $l$ far away from the soliton center ($l >> 1/m $) the
solutions $\eta_{k}(l)$ will be trigonometric functions of $kl$.
Since $E_k \approx k$ for large k , the matrix element in
(\ref{mael}) will behave as $1/k$ for large k and give rise to the
logaritmically divergent fluctuations  in eq. (\ref{fluc}). But
the charge operator with fuzzy edges $\tilde{P}_{l,d}$ defined in
(\ref{fuzzy}) has an additional integral over  $ dl'$ which will
bring down an additional convergence factor of  $1/k$ . There is
also the extra factor $ 1/d $ in the definition which goes to
zero. Altogether \beqarr \langle sol | ( \tilde{P}_{l,d} \ - \ <
\tilde{P}_{l,d}> )^2 | sol \rangle &=& \sum_{k,k'} \ | \langle
k,k' | \tilde{P}_{l,d} \ - \ < \tilde{P}_{l,d}> \ | sol \rangle |
^2 \nonumber \\
&\approx& \ {1 \over d^2} \ \sum_{k,k'} \ f(k,k') \eeqarr where
the function $f(k,k')$) has a convergent sum over  $k,k' $ .
Clearly this vanishes as $d \rightarrow \infty$. In other words,
one can define a suitable partial charge operator with fuzzy edges
for which the soliton state has a value of -1/2 with {\it no}
fluctuations ! This fractional charge then {\it is} an eigenvalue.

 It is very possible that if one were
to measure the charge in the neighborhood of the soliton's center,
one is using some such partial charge operator with fuzzy edges.
In that case the measured value of 1/2 can be promoted to the
status of an eigenvalue. Further, since these fractions are
eigenvalues of complicated partial charge operators and not of the
total charge (Number) operator Q,  there is no conceptual problem
reconciling the result with our intuition about the latter.

\subsection{Polyacetylene re-visited}

Let us describe briefly what a corresponding analysis \cite{BR}
yields for polyacetylene which is a lattice chain of some finite
number N of sites amongst which the electrons hop. I will
emphasize just the additional insights that emerge here due to the
fact that it is an even better regulated system than the field
theory in finite volume that we just studied. It has only a finite
number of degrees of freedom and hence no  ultraviolet or infrared
problems. The lattice Dirac Hamiltonian in eq(\ref{Hpoly}) yields
the following coupled equations for the electron wave functions
$b_n$ and $c_n$ at odd and even sites respectively : \beqarr
 E \ b_n \ &=& \ (1/4)c_{n+1}  \
  (\phi_{n+1} + \phi_{n} \ - \ 2/a) \ \ + \ \ (1/4) c_{n-1}
   \ (\phi_{n}+ \phi_{n-1} \ + \ 2/a) \nonumber \\
E \ c_n \ &=& \ (1/4)b_{n+1}  \
  (\phi_{n+1} + \phi_{n} \ + \ 2/a) \ \ + \ \ (1/4) b_{n-1}
   \ (\phi_{n}+ \phi_{n-1} \ - \ 2/a) \label{Deqn'}\eeqarr
Clearly, given a solution for some positive energy E we can get
another solution of negative energy (-E) by replacing $b_n
\rightarrow \ -b_n \ , \ c_n \rightarrow \ c_n \ , \ E \rightarrow
\ -E $ in the above equations. Thus positive and negative energy
solutions again come in pairs. But suppose the lattice had
altogether only an {\underline odd} number N of sites. Then the
Dirac operator can have only the same odd number of independent
solutions. Therefore there must be one or more zero-energy
solutions.

Notice that this argument for the existence of zero modes has
nothing to do with whether the background boson field is in the
topological soliton  sector or the normal vacuum sector. All you
need is an odd number of lattice sites ! One can explicitly verify
this result. Take N = 2M + 1 with, say, M odd. A zero energy
solution of the lattice Dirac equation (\ref{Deqn'}) clearly
exists in which the even site variable $c_n = 0$ for all $n$ while
the odd site variables $b_n$ satisfy
 \beq {b_{n+1} \over b_{n-1} }
\ = \ {2 \ - \ a ( \phi_n \ + \ \phi_{n-1}) \over  2 \ + \ a (
\phi_n \ + \ \phi_{n+1})} \eeq This will be true for any
background Bose field $\phi_n$. In the uniform phase (the "vacuum
sector here) $\phi_n = 1 $ for all $n$. Hence the solution is \beq
b_n \ = \ b_0  \ exp(-Kan) \ \ \ \ \ \ ;  \ \ \ \ \ \ c_n \ = \ 0
\label{zerosol}\eeq where the constant K obeys $tanh (Ka) \ = a $.
If $\phi_n$ is a soliton , say having the simple form $\phi_n \ =
\ {n \over |n|}$ then the fermion zero mode is \beq b_n \ = \ b_0
\ exp(-Ka|n|) \ \ \ \ \ \ ;  \ \ \ \ \ \ c_n \ = \ 0
\label{zerovac}\eeq Given the single zero mode in both sectors,
one can show that the ground states in both sectors will have
total charge of (-1/2) as eigenvalues.   By simply adding + 1/2 to
the definition of charge one can trivially make eigenvalues
integral in both sectors as was the case in the finite-volume
field theory. Similarly, if the number of lattice points N had
been even, there would have been either no zero modes or an even
number of them in both sectors, leading to integral total charge.

Thus the important lesson we learn is that as far as the {\bf
total} charge is concerned, there is no half integral eigenvalue
in the soliton sector {\it as compared} to the uniform phase. This
is true in finite chains even if there is a single isolated  zero
mode in the soliton sector, because a zero mode will then exist in
the vacuum sector too.

 Of course one can also define for lattice chains partial charges that pick up
the zero mode in the soliton sector but not the one in the vacuum
sector. [ Recall in the example mentioned that the zero mode in
the soliton sector given in (\ref{zerosol}) is located near the
center whereas in the vacuum sector even though a zero mode
(\ref{zerovac}) was present it was near one of the edges.] As
happened in the field theory case discussed earlier in this
section, such partial charge operators would have a half-integral
value in the soliton state.  This value would get promoted to the
status of an eigenvalue with no fluctuations if, in the limit of a
very long chain, the operator covered a region with suitably fuzzy
edges (see ref \cite{BR} for more details).

\section{Conclusion}
I have tried to outline in the last section the work that John
Bell and I had done on Fractional charge,  preceded in earlier
sections by some background on where matters stood when we began
our work.  I believe our work had thrown light on which operators do
have eigenvalues of 1/2 (as distinct from expectation values ) and
which don't. In the process we had also laid to rest the concern
that had motivated us to study this problem --- which was that the
total number operator of a physical system should not have
fractional eigenvalues. That would have made no sense. We showed
that the total charge operator defined to have integral
eigenvalues in the absence of  solitons will, in a well regulated
theory, continue to have integer eigenvalues even in the presence
of solitons. But suitable localised partial charge operators can
have fractional eigenvalues and fractional charge should be
interpreted as corresponding such an operator. Indeed these may
well be what charge measurements in experiments employ. As
mentioned already, Kivelson and Schrieffer \cite {KS} had come to
the same conclusion independently.

The work discussed above is nearly 20 years old. This article is
primarily devoted to John Bell and his work , so this is not the
place to discuss in any detail other developments on this topic
around that time or since then. However, for the sake of
completeness let me summarize in a few sentences some of these
developments .

Soon after charge 1/2 states were discovered in field theory and
polyacetylene, the  possibilities  of charge at other fractions
were unearthed both in polymer physics and in model field theories
( see for example \cite{SS} and \cite{GW}) .

A major addition to the list of fractionally charged objects
was the theoretical discovery by Laughlin in an entirely new
arena. He showed that quasi-particles in fractional quantum Hall
(FQH) systems carry fractional charge \cite{Laugh}. Subsequently
these have also been experimentally  observed in "shot-noise"
experiments \cite{shot}. To the best of my knowledge  the
quantum mechanical status of this quasi-particle charge in FQH
has not been analyzed as carefully as in the case of
polyacetylene or field theory models. The former have been
identified through ingenious but indirect arguments involving
plasma analogies and  Aharanov-Bohm/ Berry phases  rather than
by an explicit analysis of the charge operators and their
eigenstates.  The FQH  quasi-particles are more difficult states
to study. They correspond to genuinely correlated many-body
states as distinct from the polyacetylene and Jackiw-Rebbi
soliton states  considered above, which could be analysed in
terms of single particle solutions of the Dirac equation. But
broadly speaking, the status of the fractional charge in FQH is
believed to be similar to those discussed above. It is localized
near the center of the quasi-particle wavefunction and the
complementary missing fraction is believed to be near the edge
of the Hall sample.

Finally, very recently, it has been claimed by Maris \cite{He}
that some "bubbles" formed in liquid helium which  contain a
trapped electron can then fission into daughter bubbles each of
which carries a fragment of the original electron . However, in an
analysis of this phenomenon, the pioneers of fractional charge
Jackiw, Rebbi and Schrieffer have argued that these fractions in
helium bubbles are just expectation values of the sort familiar in
local charge measurements in any double well quantum system
\cite{JRS}. This is an ongoing area of work.

\section{Acknowledgements}
I would like to thank the organizers of the conference Professors
R. Bertlmann and A. Zeilinger, the University of Vienna and the
Erwin Schroedinger Institute  for their hospitality and travel
support.


\begin{references}

\bibitem{JR} R.Jackiw and C.Rebbi,  Phys. Rev.  {\bf D 13}, 3398 (1976).
\bibitem{SSH} W.P.Su, J.R.Schrieffer and A.J.Heeger, Phys. Rev. Lett {\bf 42 },1698
,(1979); Phys.Rev. {\bf B22}, 2099, (1980).
\bibitem{Rice} M.J.Rice, Phys.Lett.{\bf 71A}, 152, (1979)
\bibitem{JS}R.Jackiw and J.R.Schrieffer, Nucl.Phys . {\bf 190[FS3]}, 253, (1981)
\bibitem{Raj} R.Rajaraman, "Solitons and Instantons", North Holland , Amsterdam
(1982).
\bibitem{Kogut} J.Kogut and L.Susskind, Phys.Rev. {\bf D 11}, 395, (1975)
\bibitem{expt} I.B.Goldberg, H.R.Crowe, P.R.Newman, A.J.Heeger, and A.G.MacDiarmid,
 J.Chem.Phys. {\bf 70} 1132, (1979).
\bibitem{RB} R.Rajaraman and J.S.Bell , Phys.Lett. {\bf 116 B}, 151, (1982).
\bibitem{BR} J.S.Bell and R.Rajaraman, Nucl.Phys. {\bf B 220} [FS 8], 1 , (1983).
\bibitem{KS} S. Kivelson and J.R. Schrieffer, Phys.Rev. {\bf B 25}, 6447, (1982).
\bibitem{J et al} R.Jackiw , A.Kerman, I. Klebanov and G.Semenoff, Nucl.Phys.
{\bf B225}, [FS225], 233, (1983).
\bibitem{SS} W.P.Su and J.R.Schrieffer, Phys.Rev.Lett. {\bf 46}, 741, (1981).
\bibitem{GW} J.Goldstone and F.Wilczek, Phys.Rev.Lett. {\bf 47}, 986, (1981).
\bibitem{Laugh} R.B.Laughlin, Phys.Rev.Lett., {\bf 50}, 1395, (1983).
\bibitem {shot} R. de-Picciotto, M.Reznikov, M.Heiblum, V.Umansky,
G.Bunin, and D.Mahalu, Nature, {\bf 389}, 162 (1997)
\bibitem{He} H.Maris, J.Low Temp.Phys. {\bf 120}, 173 (2000).
\bibitem{JRS} R.Jackiw,C.Rebbi and J.R.Schrieffer, {\it "Fractional
Electrons in liquid Helium ?"}, preprint cond-mat 0012370
\end{references}
\end{document}